\documentclass{article}
\usepackage{amsmath}

 \evensidemargin0in \oddsidemargin0in \topmargin10pt
\begin{document}

\title{Relation between quantum tomography and optical Fresnel transform}
\author{Hong-Yi Fan$^{1,2}$and Li-yun Hu$^{1}\thanks{%
Corresponding author. Email: hlyun@sjtu.edu.cn}$ \\
$^{1}${\small Department of Physics, Shanghai Jiao Tong University, Shanghai
200030, China }\\
{\small \ }$^{2}${\small Department of Material Science and Engineering,
University of Science and}\\
{\small \ Technology of China, Hefei, Anhui 230026, China}}
\maketitle

\begin{abstract}
{\small Corresponding to optical Fresnel transformation characteristic of
ray transfer matrix elements }${\small (}A,B,C,D),\;AD-BC=1${\small , there
exists Fresnel operator }$F${\small (}$A,B,C,D)${\small \ in quantum optics,
we show that under the Fresnel transformation the pure position density }$%
\left\vert x\right\rangle \left\langle x\right\vert ${\small \ becomes the
tomographic density }$\left\vert x\right\rangle _{s,rs,r}\left\langle
x\right\vert ${\small , which is just the Radon transform of the Wigner
operator, i.e.}%
\begin{equation*}
F\left\vert x\right\rangle \left\langle x\right\vert F^{\dagger }=\left\vert
x\right\rangle _{s,rs,r}\left\langle x\right\vert =\int \int_{-\infty
}^{\infty }dp^{\prime }dx^{\prime }\delta \left[ x-\left( Dx^{\prime
}-Bp^{\prime }\right) \right] \Delta \left( x^{\prime },p^{\prime }\right) ,
\end{equation*}%
{\small where }$s,r${\small \ are the complex-value expression of }${\small (%
}A,B,C,D).${\small \ So the probability distribution for the Fresnel
quadrature phase is the tomography (Radon transform of Wigner function), and
the tomogram of a state }$\left\vert \psi \right\rangle ${\small \ is just
the wave function of its Fresnel transformed state }$F^{\dagger }\left\vert
\psi \right\rangle ,${\small \ i.e. }$_{s,r}\left\langle x\right\vert \left.
\psi \right\rangle =\left\langle x\right\vert F^{\dagger }\left\vert \psi
\right\rangle .${\small \ Similarly, we find }%
\begin{equation*}
F\left\vert p\right\rangle \left\langle p\right\vert F^{\dagger }=\left\vert
p\right\rangle _{s,rs,r}\left\langle p\right\vert =\int \int_{-\infty
}^{\infty }dx^{\prime }dp^{\prime }\delta \left[ p-\left( Ap^{\prime
}-Cx^{\prime }\right) \right] \Delta \left( x^{\prime },p^{\prime }\right) .
\end{equation*}
\end{abstract}


\section{Introduction}

In quantum optics theory, all possible linear combination of quadratures $X\
$and $P$\ of the oscillator field mode $a$ and $a^{\dagger }$ can be
measured by the homodyne measurement just by varying the phase of the local
oscillator. The average of the random outcomes of the measurement, at a
given local oscillator phase, is connected with the marginal distribution of
Wigner function, thus the homodyne measurement of light field permits the
reconstruction of the Wigner function of a quantum system by varying the
phase shift between two oscillators. In Ref. \cite{r1} Vogel and Risken
pointed out that the probability distribution for the rotated quadrature
phase $X_{\theta }\equiv \lbrack a^{\dagger }\exp (i\theta )+a\exp (-i\theta
)]/\sqrt{2},$ $\left[ a,a^{\dagger }\right] =1,$ which depends on only one $%
\theta $ angle, can be expressed in terms of Wigner function, and that the
reverse is also true (named as Vogel-Risken relation), i.e., one can obtain
the Wigner distribution by tomographic inversion of a set of measured
probability distributions, $P_{\theta }\left( x_{\theta }\right) ,$ of the
quadrature amplitude. Smithey, Beck and Raymer \cite{r2,r3,r4} also pointed
out that once the distribution $P_{\theta }\left( x_{\theta }\right) $ are
obtained, one can use the inverse Radon transformation familiar in
tomographic imaging to obtain the Wigner distribution and density matrix.
The Radon transform of the Wigner function is closely related to the
expectation values or densities formed with the eigenstates to the rotated
canonical observables. The field of problems of the reconstruction of the
density operator from such data is called quantum tomography. (Optical
tomographic imaging techniques derive two-dimensional data from a
three-dimensional object to obtain a slice image of the internal structure
and thus have the ability to peer inside the object noninvasively.) The
theoretical development in quantum tomography in the last decade has
progressed in the direction of determining more physical relevant parameters
of the density from tomographic data \cite{r4a,r4b,r4c}.

In \cite{r5,r6} the Radon transform of Wigner function which depends on two
continuous parameters is introduced. In this Letter we extend the rotated
quadrature phase $X_{\theta }$ to $X_{F}\equiv \left( s^{\ast }a+ra^{\dagger
}+sa^{\dagger }+r^{\ast }a\right) /\sqrt{2},$where $|s|^{2}-|r|^{2}=1,$ $%
(s,r)$ are related to a classical ray transfer matrix $\left(
\begin{array}{cc}
A & B \\
C & D%
\end{array}%
\right) $ by%
\begin{equation}
s=\frac{1}{2}\left[ A+D-i\left( B-C\right) \right] ,\;r=-\frac{1}{2}\left[
A-D+i\left( B+C\right) \right] ,\;AD-BC=1.  \label{e1}
\end{equation}%
We shall show that the $(D,B)$ related Radon transform of the Wigner
operator $\Delta \left( x,p\right) $ is just the pure state density operator
$\left\vert x\right\rangle _{s,rs,r}\left\langle x\right\vert $ formed with
the eigenstates belonging to the quadrature $x_{F}$. We name $x_{F}$ the
Fresnel transformed canonical observable, or Fresnel transformed quadrature
phase, so the probability distribution for the Fresnel quadrature phase is
the Radon transform of Wigner function. These can be expressed neatly by
\begin{eqnarray}
F\left\vert x\right\rangle \left\langle x\right\vert F^{\dagger }
&=&\left\vert x\right\rangle _{s,rs,r}\left\langle x\right\vert =\int
\int_{-\infty }^{\infty }dx^{\prime }dp^{\prime }\delta \left[ x-\left(
Dx^{\prime }-Bp^{\prime }\right) \right] \Delta \left( x^{\prime },p^{\prime
}\right) ,  \notag \\
D &=&\frac{1}{2}\left( s+s^{\ast }+r+r^{\ast }\right) ,\text{ \ }B=\frac{1}{%
2i}\left( s^{\ast }-s+r^{\ast }-r\right) ,  \label{e2}
\end{eqnarray}%
where $X_{F}\left\vert x\right\rangle _{s,r}=x\left\vert x\right\rangle
_{s,r},$ and $\left\vert x\right\rangle _{s,r}=F\left\vert x\right\rangle ,$
$\left\vert x\right\rangle $ is the eigenvector of $X_{\theta =0},$ and $F$
is the Fresnel operator corresponding to classical Fresnel transform in
optical diffraction theory which we shall describe in the following. We name
$\left\vert x\right\rangle _{s,rs,r}\left\langle x\right\vert $ the
tomographic density operator. While the $(A,C)$ related Radon transform of $%
\Delta \left( x,p\right) $ is just the pure state density operator $%
\left\vert p\right\rangle _{s,rs,r}\left\langle p\right\vert $ formed with
the eigenstates belonging to the conjugate quadrature of $X_{F},$%
\begin{eqnarray}
F\left\vert p\right\rangle \left\langle p\right\vert F^{\dagger }
&=&\left\vert p\right\rangle _{s,rs,r}\left\langle p\right\vert =\int
\int_{-\infty }^{\infty }dx^{\prime }dp^{\prime }\delta \left[ p-\left(
Dx^{\prime }-Bp^{\prime }\right) \right] \Delta \left( x^{\prime },p^{\prime
}\right) ,  \notag \\
A &=&\frac{1}{2}\left( s^{\ast }-r^{\ast }+s-r\right) ,\text{ \ }C=\frac{1}{%
2i}\left( s-r-s^{\ast }+r^{\ast }\right) .  \label{e3}
\end{eqnarray}%
Through (\ref{e2}) and (\ref{e3}) one can see how the quantum tomography is
related to optical Fresnel transform.

The optical diffraction transform is described by Fresnel integration whose
parameters $\left( A,B,C,D\right) $ are elements of a ray transfer matrix $M$
describing optical systems, $AD-BC=1,$ $M$ belongs to the unimodular
symplectic group, the input light field $f\left( x\right) $ and output light
field $g\left( x^{\prime }\right) $ are related to each other by Fresnel
integration \cite{r7,r8}
\begin{equation}
g\left( x^{\prime }\right) =\frac{1}{\sqrt{2\pi iB}}\int_{-\infty }^{\infty
}\exp \left[ \frac{i}{2B}\left( Ax^{2}-2x^{\prime }x+Dx^{\prime 2}\right) %
\right] f\left( x\right) dx.  \label{e4}
\end{equation}%
In order to find the quantum correspondence (Fresnel operator) of Fresnel
transform, the coherent state \cite{r9,r10} $\left\vert x,p\right\rangle
=\exp [-\frac{1}{4}\left( p^{2}+x^{2}\right) +\left( x+ip\right) a^{\dagger
}/\sqrt{2}]\left\vert 0\right\rangle $ is a good candidate for providing
with classical phase-space description of quantum systems, thus we construct
the following ket-bra projection operator%
\begin{equation}
\sqrt{\frac{1}{2}\left[ A+D-i\left( B-C\right) \right] }\iint\limits_{-%
\infty }^{\infty }dxdp\left\vert \left(
\begin{array}{cc}
A & B \\
C & D%
\end{array}%
\right) \left(
\begin{array}{c}
x \\
p%
\end{array}%
\right) \right\rangle \left\langle \left(
\begin{array}{c}
x \\
p%
\end{array}%
\right) \right\vert \equiv F,  \label{e5}
\end{equation}%
as the FO, where the factor $\sqrt{..}$ is attached for anticipating the
unitarity of the operator $F.$ In (\ref{e5}) the symplectic transformation $%
\left(
\begin{array}{cc}
A & B \\
C & D%
\end{array}%
\right) \left(
\begin{array}{c}
x \\
p%
\end{array}%
\right) $ mapping onto an Fresnel operator (FO) in Hilbert space is
manifestly shown through the coherent state basis. Using the notation of $%
\left\vert z\right\rangle =\exp \left[ -\frac{1}{2}|z|^{2}+za^{\dagger }%
\right] \left\vert 0\right\rangle ,$ $z=\left( x+ip\right) /\sqrt{2},$ and
introducing complex numbers
\begin{equation}
s=\frac{1}{2}\left[ A+D-i\left( B-C\right) \right] ,\;r=-\frac{1}{2}\left[
A-D+i\left( B+C\right) \right] ,\;|s|^{2}-|r|^{2}=1,  \label{e6}
\end{equation}%
such that
\begin{eqnarray}
\left\vert \left(
\begin{array}{cc}
A & B \\
C & D%
\end{array}%
\right) \left(
\begin{array}{c}
x \\
p%
\end{array}%
\right) \right\rangle &=&\left\vert \left(
\begin{array}{cc}
s & -r \\
-r^{\ast } & s^{\ast }%
\end{array}%
\right) \left(
\begin{array}{c}
z \\
z^{\ast }%
\end{array}%
\right) \right\rangle \equiv \left\vert sz-rz^{\ast }\right\rangle  \notag \\
&=&\exp \left[ -\frac{1}{2}|sz-rz^{\ast }|^{2}+(sz-rz^{\ast })a^{\dagger }%
\right] |0\rangle ,  \label{e7}
\end{eqnarray}%
so (\ref{e5}) becomes \cite{r11}%
\begin{equation}
F\left( s,r\right) =\sqrt{s}\int \frac{d^{2}z}{\pi }\left\vert sz-rz^{\ast
}\right\rangle \left\langle z\right\vert ,\;\;  \label{e8}
\end{equation}%
Using the vacuum-state projector $|0\rangle \langle 0|$ in normal ordering
of boson operators
\begin{equation}
\,\left\vert 0\right\rangle \left\langle 0\right\vert =\colon e^{-a^{\dagger
}a}\colon ,  \label{e8a}
\end{equation}%
and the technique of integration within an ordered product (IWOP) of
operators \cite{r12,r13,r14} we can directly perform the integration in (\ref%
{e8}) and obtain
\begin{eqnarray}
F\left( s,r\right) &=&\sqrt{s}\int \frac{d^{2}z}{\pi }\colon \exp \left[
-\left\vert s\right\vert ^{2}\left\vert z\right\vert ^{2}+sza^{\dagger
}+z^{\ast }\left( a-ra^{\dagger }\right) +\frac{r^{\ast }s}{2}z^{2}+\frac{%
rs^{\ast }}{2}z^{\ast 2}-a^{\dagger }a\right] \colon  \notag \\
&=&\frac{1}{\sqrt{s^{\ast }}}\exp \left( -\frac{r}{2s^{\ast }}a^{\dagger
2}\right) \colon \exp \left\{ \left( \frac{1}{s^{\ast }}-1\right) a^{\dagger
}a\right\} \colon \exp \left( \frac{r^{\ast }}{2s^{\ast }}a^{2}\right) ,
\label{e9}
\end{eqnarray}%
Note that this can be identified as a generalized squeezing operator with
three real parameters \cite{r15,r16,r17}. It then follows
\begin{equation}
\left\langle z\right\vert F\left( s,r\right) \left\vert z^{\prime
}\right\rangle =\frac{1}{\sqrt{s^{\ast }}}\exp \left[ -\frac{\left\vert
z\right\vert ^{2}}{2}-\frac{\left\vert z^{\prime }\right\vert ^{2}}{2}-\frac{%
r}{2s^{\ast }}z^{\ast 2}-\frac{r^{\ast }}{2s^{\ast }}z^{\prime 2}+\frac{%
z^{\ast }z^{\prime }}{s^{\ast }}\right] .  \label{e10}
\end{equation}%
Then using the overlap between the coordinate eigenvector and the coherent
state
\begin{equation}
\left\langle x\right\vert \left. z\right\rangle =\pi ^{-1/4}\exp \left( -%
\frac{x^{2}}{2}+\sqrt{2}xz-\frac{z^{2}}{2}-\frac{\left\vert z\right\vert ^{2}%
}{2}\right) ,  \label{e11}
\end{equation}%
and the completeness relation of coherent state we obtain the matrix element
of $F\left( s,r\right) $ in coordinate representation,
\begin{eqnarray}
\left\langle x^{\prime }\right\vert F\left( s,r\right) \left\vert
x\right\rangle &=&\int \frac{d^{2}z}{\pi }\left\langle x^{\prime
}\right\vert \left. z\right\rangle \left\langle z\right\vert F_{1}\left(
s,r\right) \int \frac{d^{2}z^{\prime }}{\pi }\left\vert z^{\prime
}\right\rangle \left\langle z^{\prime }\right\vert \left. x\right\rangle
\notag \\
&=&\frac{1}{\sqrt{2\pi iB}}\exp \left[ \frac{i}{2B}\left( Ax^{2}-2x^{\prime
}x+Dx^{\prime 2}\right) \right] ,  \label{e12}
\end{eqnarray}%
which is just the kernel of generalized Fresnel transform in (\ref{e4}).

Now if we define $g\left( x^{\prime }\right) =\left\langle x^{\prime
}\right\vert \left. g\right\rangle $, $f\left( x\right) =\left\langle
x\right\vert \left. f\right\rangle $ and using Eq. (\ref{e12}), we can
rewrite Fresnel transform in Eq. (\ref{e4}) as%
\begin{equation}
\left\langle x^{\prime }\right\vert \left. g\right\rangle =\int_{-\infty
}^{\infty }dx\left\langle x^{\prime }\right\vert F\left( A,B,C\right)
\left\vert x\right\rangle \left\langle x\right\vert \left. f\right\rangle
=\left\langle x^{\prime }\right\vert F\left( A,B,C\right) \left\vert
f\right\rangle .  \label{e13}
\end{equation}%
Therefore, the 1-dimensional FT in classical optics corresponds to the
1-mode FO $F\left( A,B,C\right) $ operating on state vector $\left\vert
f\right\rangle $ in Hilbert space, i.e. $\left\vert g\right\rangle =F\left(
A,B,C\right) \left\vert f\right\rangle $. Using the Fock representation of $%
\left\vert x\right\rangle ,$
\begin{equation}
\left\vert x\right\rangle =\pi ^{-1/4}\exp \left\{ -\frac{1}{2}x^{2}+\sqrt{2}%
xa^{\dagger }-\frac{a^{\dagger 2}}{2}\right\} |0\rangle ,  \label{e14}
\end{equation}%
and (\ref{e8}) as well as (\ref{e11}) we calculate%
\begin{eqnarray}
F\left( s,r\right) \left\vert x\right\rangle &=&\sqrt{s}\int \frac{d^{2}z}{%
\pi }\left\vert sz-rz^{\ast }\right\rangle \left\langle z\right\vert \left.
x\right\rangle  \notag \\
&=&\pi ^{-1/4}\sqrt{s}\int \frac{d^{2}z}{\pi }\exp \left[ -\frac{1}{2}%
|sz-rz^{\ast }|^{2}+(sz-rz^{\ast })a^{\dagger }\right] |0\rangle  \notag \\
&&\exp \left( -\frac{x^{2}}{2}+\sqrt{2}xz^{\ast }-\frac{z^{\ast 2}}{2}-\frac{%
\left\vert z\right\vert ^{2}}{2}\right) =\left\vert x\right\rangle _{s,r},
\label{e15}
\end{eqnarray}%
where%
\begin{equation}
\left\vert x\right\rangle _{s,r}\equiv \frac{\pi ^{-1/4}}{\sqrt{s^{\ast
}+\allowbreak r^{\ast }}}\exp \{-\frac{s^{\ast }-r^{\ast }}{s^{\ast
}+r^{\ast }}\frac{x^{2}}{2}+\frac{\sqrt{2}x}{s^{\ast }+r^{\ast }}a^{\dagger
}-\frac{s+r}{s^{\ast }+r^{\ast }}\frac{a^{\dagger 2}}{2}\}\left\vert
0\right\rangle .  \label{e16}
\end{equation}%
We can see that $\left\vert x\right\rangle _{s,r}$ $=F\left( s,r\right)
\left\vert x\right\rangle $ is the eigenstate of Fresnel transformed
quadrature phase, because from%
\begin{equation}
a\left\vert x\right\rangle _{s,r}=\left( \frac{\sqrt{2}x}{s^{\ast }+r^{\ast }%
}-\frac{s+r}{s^{\ast }+r^{\ast }}a^{\dagger }\right) \left\vert
x\right\rangle _{s,r},  \label{e17}
\end{equation}%
so
\begin{equation}
\left[ \left( s^{\ast }+r^{\ast }\right) a+\left( s+r\right) a^{\dagger }%
\right] \left\vert x\right\rangle _{s,r}=\sqrt{2}x\left\vert x\right\rangle
_{s,r}.  \label{e18}
\end{equation}%
This can be further confirmed by examining its completeness relation. Using
the IWOP technique and (\ref{e8a}) as well as $\frac{1}{\left\vert
s+r\right\vert ^{2}}=\frac{s^{\ast }-r^{\ast }}{2\left( r^{\ast }+s^{\ast
}\right) }+\frac{s-r}{2\left( r+s\right) },$ we can prove that $\left\vert
x\right\rangle _{s,r}$ make up a complete set (named as the tomography
representation),%
\begin{eqnarray}
\int_{-\infty }^{\infty }dx\left\vert x\right\rangle
_{s,r}{}_{s,r}\left\langle x\right\vert &=&\int_{-\infty }^{\infty }\frac{dx%
}{\left\vert s+r\right\vert \sqrt{\pi }}\colon \exp \left\{ -\left( \frac{%
s^{\ast }-r^{\ast }}{s^{\ast }+r^{\ast }}+\frac{s-r}{s+r}\right) \frac{x^{2}%
}{2}-a^{\dagger }a\right.  \notag \\
&&\left. +\sqrt{2}x\left( \frac{a^{\dagger }}{s^{\ast }+r^{\ast }}+\frac{a}{%
s+r}\right) -\frac{s+r}{s^{\ast }+r^{\ast }}\frac{a^{\dagger 2}}{2}-\frac{%
s^{\ast }+r^{\ast }}{s+r}\frac{a^{2}}{2}\right\} \colon  \notag \\
&=&\int_{-\infty }^{\infty }\frac{dx}{\left\vert s+r\right\vert \sqrt{\pi }}%
\colon \exp \left\{ -\frac{1}{\left\vert s+r\right\vert ^{2}}\left( x-\frac{%
s^{\ast }a+ra^{\dagger }+sa^{\dagger }+r^{\ast }a}{\sqrt{2}}\right)
^{2}\right\} \colon =1,  \label{e19}
\end{eqnarray}%
then using (\ref{e6}) and $X=\frac{a+a^{\dagger }}{\sqrt{2}},P=i\frac{%
a^{\dagger }-a}{\sqrt{2}}$ as well as $s^{\ast }+r^{\ast }=D+iB,$\ \ $%
s^{\ast }-r^{\ast }=A-iC,$ we can reform (\ref{e19}) as%
\begin{equation}
1=\frac{1}{\sqrt{D^{2}+B^{2}}}\int_{-\infty }^{\infty }\frac{dx}{\sqrt{\pi }}%
\colon \exp \left\{ -\frac{1}{D^{2}+B^{2}}\left( x-DX+BP\right) ^{2}\right\}
\colon ,  \label{e20}
\end{equation}%
and $\left\vert x\right\rangle _{s,r}$ is express as%
\begin{equation}
\left\vert x\right\rangle _{s,r}=\frac{\pi ^{-1/4}}{\sqrt{D+iB}}\exp \left\{
-\frac{A-iC}{D+iB}\frac{x^{2}}{2}+\frac{\sqrt{2}x}{D+iB}a^{\dagger }-\frac{%
D-iB}{D+iB}\frac{a^{\dagger 2}}{2}\right\} \left\vert 0\right\rangle .
\label{e22}
\end{equation}%
and (\ref{e18}) becomes%
\begin{equation}
\left( DX-BP\right) \left\vert x\right\rangle _{s,r}=x\left\vert
x\right\rangle _{s,r}.  \label{e21}
\end{equation}%
According to the Weyl quantization scheme \cite{r18}%
\begin{equation}
H\left( X,P\right) =\int \int_{-\infty }^{\infty }dpdx\Delta \left(
x,p\right) h\left( x,p\right) ,  \label{17}
\end{equation}%
where $h\left( x,p\right) $ is the Weyl correspondence of $H\left(
X,P\right) ,$ $\Delta \left( x,p\right) $ is the Wigner operator \cite%
{r19,r20},%
\begin{equation}
\Delta \left( x,p\right) =\frac{1}{2\pi }\int_{-\infty }^{\infty
}due^{ipu}\left\vert x+\frac{u}{2}\right\rangle \left\langle x-\frac{u}{2}%
\right\vert .  \label{18}
\end{equation}%
we know that the classical Weyl correspondence of the projection operator $%
\left\vert x\right\rangle _{s,rs,r}\left\langle x\right\vert $ is
\begin{eqnarray}
2\pi Tr\left[ \Delta \left( x^{\prime },p^{\prime }\right) \left\vert
x\right\rangle _{s,rs,r}\left\langle x\right\vert \right] &=&\left.
_{s,r}\left\langle x\right\vert \right. \int_{-\infty }^{\infty
}due^{ip^{\prime }u}\left\vert x^{\prime }+\frac{u}{2}\right\rangle
\left\langle x^{\prime }-\frac{u}{2}\right\vert \left. x\right\rangle _{s,r}
\notag \\
&=&\frac{1}{2\pi B}\int_{-\infty }^{\infty }du\exp \left[ ip^{\prime }u+%
\frac{i}{B}u\left( x-Dx^{\prime }\right) \right]  \notag \\
&=&\delta \left[ x-\left( Dx^{\prime }-Bp^{\prime }\right) \right] .
\label{19}
\end{eqnarray}%
From we see that under the Fresnel transformation (\ref{e15}) the pure
position density $\left\vert x\right\rangle \left\langle x\right\vert $
becomes the tomographic density $\left\vert x\right\rangle
_{s,rs,r}\left\langle x\right\vert $, and further from (\ref{17}) and (\ref%
{19}) we see that it is just the Radon transform of the Wigner operator, i.e.%
\begin{equation}
F\left\vert x\right\rangle \left\langle x\right\vert F^{\dagger }=\left\vert
x\right\rangle _{s,rs,r}\left\langle x\right\vert =\int \int_{-\infty
}^{\infty }dx^{\prime }dp^{\prime }\delta \left[ x-\left( Dx^{\prime
}-Bp^{\prime }\right) \right] \Delta \left( x^{\prime },p^{\prime }\right) .
\label{20}
\end{equation}%
Therefore, the probability distribution for the Fresnel quadrature phase is
the Radon transform of Wigner function%
\begin{equation}
|\left\langle x\right\vert F^{\dagger }\left\vert \psi \right\rangle
|^{2}=|_{s,r}\left\langle x\right\vert \left. \psi \right\rangle |^{2}=\int
\int_{-\infty }^{\infty }dx^{\prime }dp^{\prime }\delta \left[ x-\left(
Dx^{\prime }-Bp^{\prime }\right) \right] \left\langle \psi \right\vert
\Delta \left( x^{\prime },p^{\prime }\right) \left\vert \psi \right\rangle .
\label{21}
\end{equation}%
Moreover, the tomogram of quantum state $\left\vert \psi \right\rangle $ is
just the wave function of its Fresnel transformed state $F^{\dagger
}\left\vert \psi \right\rangle ,$ i.e. $_{s,r}\left\langle x\right\vert
\left. \psi \right\rangle =\left\langle x\right\vert F^{\dagger }\left\vert
\psi \right\rangle .$ (\ref{20}) and (\ref{21}) are the new relation between
quantum tomography and optical Fresnel transform, which may provide
experimentalists to figure out new approach for generating tomography.

Similarly, we find that for momentum density,
\begin{equation}
F\left\vert p\right\rangle \left\langle p\right\vert F^{\dagger }=\left\vert
p\right\rangle _{s,rs,r}\left\langle p\right\vert =\int \int_{-\infty
}^{\infty }dx^{\prime }dp^{\prime }\delta \left[ p-\left( Ap^{\prime
}-Cx^{\prime }\right) \right] \Delta \left( x^{\prime },p^{\prime }\right) ,
\label{22}
\end{equation}%
where
\begin{equation}
F\left\vert p\right\rangle =\left\vert p\right\rangle _{s,r}=\frac{\pi
^{-1/4}}{\sqrt{A-iC}}\exp \left\{ -\frac{D+iB}{A-iC}\frac{p^{2}}{2}+\frac{%
\sqrt{2}ip}{A-iC}a^{\dagger }+\frac{A+iC}{A-iC}\frac{a^{\dagger 2}}{2}%
\right\} \left\vert 0\right\rangle .  \label{23}
\end{equation}

As an application of the relation (\ref{20}), we recall that the Fresnel
operator $F(r,s)$ makes up a loyal representation of the symplectic group,
i.e.%
\begin{align}
& F(r,s)F^{\prime }(r^{\prime },s^{\prime })=\sqrt{ss^{\prime }}\int \frac{%
d^{2}z}{\pi }\int \frac{d^{2}z^{\prime }}{\pi }\left\vert sz-rz^{\ast
}\right\rangle \left\langle z\right. \left\vert s^{\prime }z^{\prime
}-r^{\prime }z^{\prime \ast }\right\rangle \left\langle z^{\prime
}\right\vert  \notag \\
& =\frac{1}{\sqrt{s^{\prime \prime \ast }}}\exp \left( -\frac{r^{\prime
\prime }}{2s^{\prime \prime \ast }}a^{\dagger 2}\right) \colon \exp \left\{
\left( \frac{1}{s^{\prime \prime \ast }}-1\right) a^{\dagger }a\right\}
\colon \exp \left( \frac{r^{\prime \prime \ast }}{2s^{\prime \prime \ast }}%
a^{2}\right) =F(r^{\prime \prime },s^{\prime \prime }),  \label{5}
\end{align}%
where%
\begin{equation}
\left(
\begin{array}{cc}
s^{\prime \prime } & r^{\prime \prime } \\
-r^{\prime \prime \ast } & s^{\prime \prime \ast }%
\end{array}%
\right) =\left(
\begin{array}{cc}
s & r \\
-r^{\ast } & s^{\ast }%
\end{array}%
\right) \left(
\begin{array}{cc}
s^{\prime } & r^{\prime } \\
-r^{\prime \ast } & s^{\prime \ast }%
\end{array}%
\right) ,\text{ }\left\vert s^{\prime \prime }\right\vert ^{2}-\left\vert
r^{\prime \prime }\right\vert ^{2}=1,  \label{6}
\end{equation}%
or%
\begin{equation}
\left(
\begin{array}{cc}
A^{\prime \prime } & B^{\prime \prime } \\
C^{\prime \prime } & D^{\prime \prime }%
\end{array}%
\right) =\left(
\begin{array}{cc}
A & B \\
C & D%
\end{array}%
\right) \left(
\begin{array}{cc}
A^{\prime } & B^{\prime } \\
C^{\prime } & D^{\prime }%
\end{array}%
\right) =\left(
\begin{array}{cc}
AA^{\prime }+BC^{\prime } & AB^{\prime }+BD^{\prime } \\
A^{\prime }C+C^{\prime }D & B^{\prime }C+DD^{\prime }%
\end{array}%
\right) .  \label{8}
\end{equation}%
It then follows from (\ref{20}) (\ref{8}) and that
\begin{eqnarray}
&&F^{\prime }(r^{\prime },s^{\prime })F(r,s)\left\vert x\right\rangle
\left\langle x\right\vert F^{\dagger }(r,s)F^{\prime \dagger }(r^{\prime
},s^{\prime })  \notag \\
&=&\int \int_{-\infty }^{\infty }dx^{\prime }dp^{\prime }\delta \left[
x-\left( \left( B^{\prime }C+DD^{\prime }\right) x^{\prime }-\left(
AB^{\prime }+BD^{\prime }\right) p^{\prime }\right) \right] \Delta \left(
x^{\prime },p^{\prime }\right)  \notag \\
&=&\int \int_{-\infty }^{\infty }dx^{\prime }dp^{\prime }\delta \left[
x-\left( D^{\prime \prime }x^{\prime }-B^{\prime \prime }p^{\prime }\right) %
\right] \Delta \left( x^{\prime },p^{\prime }\right) ,  \label{7}
\end{eqnarray}%
in this way the complicated Radon transform of tomography can be viewed as
the sequential operation of two Fresnel transforms.

In sum, we have revealed the new relation connecting optical Fresnel
transformation to Radon transformation in quantum tomography{\small ,} i.e.
the probability distribution for the Fresnel quadrature phase is the
tomography (Radon transform of Wigner function). The tomography
representation $_{s,r}\left\langle x\right\vert $ is set up, based on which
the tomogram of a state $\left\vert \psi \right\rangle $\ is just the wave
function of its Fresnel transformed state $F^{\dagger }\left\vert \psi
\right\rangle ,$\ i.e. $_{s,r}\left\langle x\right\vert \left. \psi
\right\rangle =\left\langle x\right\vert F^{\dagger }\left\vert \psi
\right\rangle .$ The group property of Fresnel operators help us to analyze
complicated Radon transforms in terms of some sequential Fresnel
transformations. The new relation may provide experimentalists to figure out
new approach for realizing tomography.

We would like to acknowledge support from the National Natural Science
Foundation of China under Grant Nos. 10775097 and 10475056.

\end{document}